\documentclass[%
 ,twocolumn%
 ,secnumarabic%
,amssymb, amsmath,nobibnotes, aps, prl]{revtex4-1}
\usepackage{bm}%

\usepackage{graphicx}
\usepackage{epstopdf}

\newcommand\err{\varepsilon_{rr}}
\newcommand\eqq{\varepsilon_{\theta\theta}}
\newcommand\Rin{R_{in}}
\renewcommand\b{\tau}

\begin{document}

\title{Far From Threshold Buckling Analysis of   Thin Films}%
\author{B. Davidovitch$^{1}$, R.D. Schroll$^{1}$,  D.~Vella$^{2,3}$, M.~Adda-Bedia$^{2}$, E.~Cerda$^{4}$}

\affiliation{
$^1$
Physics Department, University of Massachusetts, Amherst MA 01003\\
$^2$Laboratoire de Physique Statistique, Ecole Normale Sup\'erieure, UPMC Paris 06, Universit\'e Paris Diderot, CNRS, 24 rue Lhomond, 75005 Paris, France\\
$^3$ITG, Department of Applied Mathematics and Theoretical Physics, University of Cambridge, Wilberforce Road, Cambridge, CB3 0WA, UK\\
$^4$Departamento de F\'{i}sica, Universidad de Santiago, Av.~Ecuador 3493, Santiago, Chile}

%
\date{\today}%

\begin{abstract}
Thin films buckle easily and form wrinkled states in regions of well defined  size. The extent of a wrinkled region is typically assumed to reflect the zone of  in-plane compressive stresses prior to buckling, but recent experiments on ultrathin sheets have shown that wrinkling patterns are significantly longer and follow different scaling laws than those predicted by standard buckling theory. Here we focus on a simple setup to show the striking differences between near-threshold buckling and the analysis of wrinkle patterns in very thin films, which are typically far from threshold.

\end{abstract}

\maketitle

The growing interest in developing technologies at smaller and smaller scales has posed new questions and challenges for scientists to understand the mechanical behavior of tiny structures. Engineered  films  with thickness ranging from nano to microscales and designed  for different applications are among the ubiquitous examples of flexible structures  that buckle under very small loads. More interestingly, these buckling instabilities usually  develop into  wrinkled patterns that form a dramatic display of the applied stress field \cite{Hutchinson,Groenewold}. Wrinkles align perpendicularly to the compression direction, depicting the principal lines of stress and providing through their geometry new tools for mechanical characterization.

  Buckling theory is regularly used  to understand these patterns in macroscopic plates when the deformations are small perturbations of the initial flat state.  However,  it has been known since Wagner \cite{Wagner,Mansfield} that  plates buckled under loads well in excess of those necessary to initiate buckling show an asymptotic state different from the one observed near threshold. The stress nearly vanishes in the compression direction and the plate acquires fine wrinkles that mark the region where the compressive stress has collapsed. This asymptotic state is unusual in macroscopic plates, but very likely to happen in very thin films, since their threshold load values are very small.



 A better insight into this ``collapsed" wrinkled state was provided by the recent discovery of scaling relations  between wrinkle {\emph{wavelength}, film thickness and applied tension in stretched films of rectangular shape \cite{Cerda01}. This theory, and later applications \cite{Cerda05}, assumed that the wrinkle {\emph{length}}  is determined by the in-plane compressive region prior to buckling. However, recent experiments and theoretical work show that the length of wrinkles in very thin films is significantly larger than predictions based on the stress field near the onset of buckling \cite{Huang,Vella}, and thus indicate that our conceptual understanding of the far-from-threshold wrinkled state is still lacking. In this Letter we present a novel analysis of the far-from-threshold limit and predict a new scaling law for the extent of the wrinkled region in very thin sheets. Details of this asymptotic theory will be published elsewhere \cite{Davidovitch10}.

 \begin{figure}[h]
 \center\includegraphics[width= 6cm]{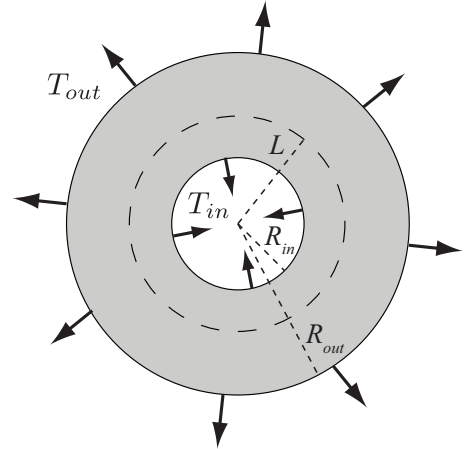}
 \caption{The classical Lam\'e configuration where a mismatch between the inner and outer stresses produces a compressive region for $r<L$.}
 \label{fig_1}
 \end{figure}

 In order to study the essential differences between the near-threshold (NT) and far-from-threshold (FFT) regimes, we focus here on the circular configuration shown in Fig.~\ref{fig_1} where an annular film of inner radius $\Rin$ and outer radius $R_{out}$ is stretched differentially by radial forces per length $T_{out}$ at $r=R_{out}$ and $T_{in}>T_{out}$ at $r=\Rin$. Similar geometries have been used   to study the wrinkling pattern under different types of central loads, such as the impact of fast projectiles \cite{vermorel09}, the deadhesion and wrinkling of a thin sheet loaded at a point \cite{chopin08}, and the wrinkling and folding of floating membranes \cite{Huang, holmes10, menon10}. However, Fig.~\ref{fig_1} exhibits the simplest load distribution leading to wrinkling with variable length and wavelength. It is a classical problem of linear elasticity \cite{Timoshenko} to obtain the radial and hoop stresses of the planar, axially-symmetric state from radial force balance: 
 \begin{eqnarray}
 &&\frac{1}{r}\frac{\partial}{\partial r}(r \sigma_{rr}) - \frac{\sigma_{\theta\theta}}{r} = 0 \ ,
 \label{2ndFvK} \\
 \text{where}  \ \ &&\sigma_{ii} = \frac{Y}{1-\nu^2}(\varepsilon_{ii} + \nu \varepsilon_{jj})   \  , \ (i,j) = (r,\theta)
 \label{stressstrain}
 \end{eqnarray}
 The radial and hoop strains are, respectively,
 $\err = \frac{\partial u_r}{\partial r},  \ \eqq = \frac{u_r}{r}$,
 and $u_r(r)$ is the only nonvanishing displacement. Here, $Y\!=\!Et$ is the stretching modulus, $E$ is the Young modulus, $t$ the film thickness, and $\nu$ is the Poisson ratio. Focussing for simplicity on the limit $R_{out} \gg \Rin$, the solution is:
\begin{equation}
\sigma_{rr}=(T_{out}+\Delta T \,\frac{\Rin^2}{r^2}) \  ; \ \sigma_{\theta\theta}=(T_{out}-\Delta T \, \frac{\Rin^2}{r^2})
\label{Lame1}
\end{equation}
where $\Delta T=T_{in}-T_{out}$. 
This solution is usually named after Timoshenko, although it was first studied by Lam\'e \cite{Timoshenko}.
In this solution the radial stress is tensile everywhere, but if the stress ratio $T_{in}/T_{out} >2$ the hoop stress becomes compressive for $\Rin\!<\!r\!<\!L_{\mbox{\tiny NT}}$. Here:
 \begin{eqnarray}
L_{\mbox{\tiny NT}}=\Rin \, \sqrt{T_{in}/T_{out}-1} \approx \Rin\, (T_{in}/T_{out})^{1/2} \ .
\label{Case1}
\end{eqnarray}
The existence of compression leads to the formation of a buckled state if
the film is sufficiently thin that it becomes energetically favorable to relieve the compressive stress by bending.
The Lam\'e problem is thus characterized by two independent dimensionless groups:
\begin{equation}
\b\equiv \frac{T_{in}}{T_{out}} \ \ ; \ \ \epsilon \equiv \frac{B}{\Rin^2T_{out}}
\label{dimnensionless}
\end{equation}
where $B = Et^2/12(1-\nu^2)$ is the bending modulus. Buckling is expected for a fixed value of $\b>2$ when $\epsilon$ is smaller than a critical value $\epsilon_c$.
Our analysis assumes reducing $\epsilon$ from NT ($\epsilon \lesssim \epsilon_c$) to FFT conditions ($\epsilon \ll 1$) by, for example, varying the film thickness $t$.

Standard buckling theory in the NT regime consists of a stability analysis of the $1^{st}$ F\"oppl\textendash Von K\'arm\'an (FvK) equation \cite{Mansfield}:
\begin{equation}
B \left(\frac{d^2}{dr^2} - \frac{m^2}{r^2}\right)^2 f = - \sigma_{\theta\theta} \frac{m^2}{r^2} f + \sigma_{rr} \frac{d^2f}{dr^2}
\label{FvK22}
\end{equation}
where the out-of-plane displacement is assumed to be of the form $\zeta = f(r)\cos(m\theta)$, and $\sigma_{rr},\sigma_{\theta\theta}$ assume their Lam\'e form~(\ref{Lame1}). While the energy and wrinkled extent in the NT regime are determined by Eqs.~(\ref{Lame1},\ref{Case1}), bifurcation analysis (assuming $f(r)$ is infinitesimal) yields the numerical value of $\epsilon_c$ and the exact number of wrinkles $m$ of the emerging pattern \cite{Adams}. Simple dimensional analysis implies the general result $m = g(\tau,\epsilon)$, where $g$ is some dimensionless function.
Near threshold, the Lam\'e solution assures that $\sigma_{rr}\sim\sigma_{\theta\theta}$. If $T_{out} \! \ll \! T_{in}$ we expect $m \!\gg\!1$ and one notices that the out-of-plane stretching force in the hoop direction is much larger than in the radial direction.
Thus, the bending forces on the LHS of Eq.~(\ref{FvK22}) are balanced by the out-of-plane hoop stretching on the RHS, and one obtains the NT scaling law:
\begin{equation}
m \sim \tilde{g}(\b) \Rin \sqrt{T_{out}/B} \ ,
\label{NT:mscale}
\end{equation}
for some $\tilde{g}(\b)$.
This linear relation between the number of wrinkles and hole radius was confirmed in Ref.~\cite{geminard04}  in experiments with  macroscopic rubber membranes, which are expected to be under NT conditions. Similarly, Refs.~\cite{geminard04, Cerda05}  reported that for large samples the wrinkle extent $L$ increases as the square root of the inner tension, in agreement with Eq.~(\ref{Case1}).

Let us turn now to our main focus: the FFT regime $\epsilon \! \ll \! 1$. Motivated by experiments \cite{Huang} and following Wagner's ideas, we assume that the sheet is composed of two parts: a wrinkled region in $\Rin<r\!<\!L$ with a collapsed hoop stress $\sigma_{\theta\theta} \to 0$, and an outer annulus $L\!<\!r\!<\!R_{out}$ in which the sheet remains planar with stresses following the Lam\'{e} form~(\ref{Lame1}) appropriately modified. We shall prove below that for $\epsilon \!\ll\! 1$ a state with $L\!>\!\Rin$ is energetically favorable to the Lam\'e state, which corresponds to $L\!=\!\Rin$. Thus, wrinkling  is a mechanism for releasing elastic energy in the film.
For $\Rin\!<\!r\!<\!L$, Eqs.~(\ref{2ndFvK},\ref{stressstrain}) yield:
\begin{equation}
\sigma_{rr}=T_{in}\frac{\Rin}{r} \  ;  \ \err =(T_{in}/Y)\frac{\Rin}{r} \  ;  \ \eqq = -\nu \err \ .
\label{FFT1}
\end{equation}
 Since the radial stress must be continuous we find that the stresses in the outer annulus have the Lam\'e form~(\ref{Lame1}) with $\Rin \!\to\! L$ and $\Delta T \!\to\! \Delta T(L) \!=\!   T_{in}\Rin/L \!-\!T_{out}$. For a given wrinkle extent $L$, the FFT stresses are now fully characterized by Eqs.~(\ref{Lame1},\ref{FFT1}).
 Moreover, the radial displacement $u_r(L)$ must also be continuous, and since the Lam\'e solution at $r \geq L$ implies a link between $u_r(L)$ and $\sigma_{rr}(L)$ one obtains the radial displacement at $r\!<\!L$ \cite{Davidovitch10}:
 \begin{eqnarray}
u_r(r) &=&\Rin\, (T_{in}/Y)\log\left(\frac{r}{L}\right) + u_r(L) \ , \ \ \text{with} \nonumber \\
u_r(L) &=& [2 L\,(T_{out}/Y)- (1+\nu)\Rin\,(T_{in}/Y)].
 \label{ur1}
 \end{eqnarray}
We shall determine the actual wrinkle extent by minimizing the energy over all allowed values of $L$.

 Before turning to energy calculations, let us highlight some important aspects of the FFT solution. First, as
 Eq.~(\ref{FFT1}) indicates, there is a pure traction along the radial direction producing a Poisson effect   $\eqq=-\nu \err<0$ in the azimuthal direction and reducing the perimeter at radius $r$ by a total length $-2\pi r \nu T_{in}/Y$.
 This local contraction shows that for $\nu\neq0$ the film is not inextensible in the  azimuthal direction as was assumed in \cite{Cerda01}. However, a similar  constraint arises: there is an excess in length when this contraction is not compatible with the geometrical shortening of the perimeter length $2\pi u_r(r)$ that is generated by the inwardly radial displacement $u_r<0$.
 In order to reduce stretching energy, this excess of length is relieved by out-of-plane displacement that is highly oscillatory in the azimuthal direction. Using the  weakly nonlinear strain-displacement relations \cite{Mansfield} and the relation $\eqq=-\nu \err$, we obtain
 \begin{eqnarray}
 \int_0^{2\pi}\!\!r d\theta \ \left[\frac{u_r}{r}+\frac{1}{2r^2}\left(\frac{\partial \zeta}{\partial \theta}\right)^2 \right]= -\int_0^{2\pi} \! r d\theta  \ \nu \err.
 \label{geometryrelation}
 \end{eqnarray}
 Assuming again that $\zeta = f(r)\cos(m\theta)$ one finds:
 \begin{equation}
\frac{1}{4}\frac{m^2f^2}{r^2}=- \frac{u_r}{r} -\nu \err
\label{FFThoopstrain}
\end{equation}
The positivity of the LHS of Eq.~(\ref{FFThoopstrain}) 
manifests that out of plane deformation is possible only if there is an excess of length of the circle perimeter. An algebraic manipulation of Eqs.~(\ref{ur1},\ref{FFThoopstrain}) shows that the RHS is positive for $r<L$ only if $L \!\leq \!\Rin\, T_{in}/ 2 T_{out}$, thus providing an upper bound for the wrinkled extent.
Additionally, Eq.~(\ref{FFThoopstrain}) indicates that the product $m f(r)$ must remain finite in the FFT limit, in contrast to the NT regime where $f(r)$ is infinitesimal. Indeed, our FFT analysis is based on an asymptotic series in which the expansion parameter is $\sqrt{\epsilon}$, in contrast to the NT regime, where the small parameter is the distance to threshold $\epsilon\!-\!\epsilon_c$ \cite{Davidovitch10}.

In order to determine the wrinkled extent, $L$, we compute the elastic energy $U_E$ of the FFT stress field, where
\begin{equation}
U_E=\frac{1}{2}\int_A \ dA \ (\sigma_{rr}\err+\sigma_{\theta\theta}\eqq )\ .
\label{elasticenergy}
\end{equation}
A straightforward calculation using Eqs.~(\ref{Lame1},\ref{FFT1}) yields:
\begin{eqnarray}
&&U_E=(\pi/Y)\left\{(1-\nu)(R_{out} T_{out})^2+(\Rin\, T_{in})^2\ln (L/\Rin)\right.
\nonumber\\
&&+\left. 2(L\, T_{out} - \Rin\,T_{in})^2-(1-\nu) (\Rin\, T_{in})^2 +{\cal O}(1/R_{out}^2)\right\}
\nonumber\\
\label{U0}
\end{eqnarray}
Note that this energy does not include the costs of bending and out-of-plane stretching of the membrane. These may be shown to be higher order contributions in $\epsilon$ in the FFT limit \cite{Davidovitch10}, but are nevertheless crucial for obtaining the exact limit of the number of wrinkles, $m$. Since our problem involves applied constant forces at $r=\Rin,R_{out}$, we must minimize the mechanical energy $U\!=\! U_E \!-\! W$, where $W \!=\! 2\pi [T_{out}R_{out}\,u_r(R_{out}) \!-\! T_{in}\Rin\,u_r(\Rin)]$ is the exerted work.
Minimizing $U$ as a function of wrinkle extent is analogous to fracture mechanics problems, in which one minimizes the mechanical energy as a function of crack length under constant load conditions \cite{Lawn}. Like cracks, wrinkles provide a route for the release of elastic energy. Using a general result from elasticity theory for bodies under constant external loads~\cite{Lawn}, we find that $W\!=\!2U_E$, and hence $U\!=\! -U_E$. In order to minimize $U$ we notice that its first derivative $\partial_LU=-\pi(2 L\,  T_{out}-\Rin\, T_{in} )^2/(YL)\leq 0$ is zero for:
\begin{equation}
L_{\mbox{\tiny FFT}} = \frac{\Rin}{2}\frac{T_{in}}{T_{out}} \ .
\label{wrinkleslength}
\end{equation}
Interestingly, this result is identical to the upper bound that we found above, and also assures continuity of the hoop stress $\sigma_{\theta\theta}(L_{\mbox{\tiny FFT}})$. Thus, energy minimization
naturally yields a value for the stress at the tip of the wrinkles that smoothly matches the flat region in the film to the highly wrinkled one.
The quadratic form of the first derivative and Fig.~2 show that Eq.~(\ref{wrinkleslength}) corresponds to an inflection point of $U$, but the upper bound guarantees that this is the actual length in the FFT limit. Fig.~2 also indicates that at the FFT limit, where corrections to the mechanical energy $U$ are $\sim \sqrt{\epsilon}$, the  energy at the inflection point $U(L_{\mbox{\tiny FFT}})$ is lower than the Lam\'e value $U(\Rin)$.
Thus the energy is lowered from the NT (Lam\'e) value to the FFT asymptotic limit, as $\epsilon$ is reduced from its critical value $\epsilon_c$ to $\epsilon \! \ll \! 1$.
Equation (\ref{wrinkleslength}) is the central result of our analysis. It reflects a linear scaling of the extent of the wrinkles in the FFT regime with the ratio $T_{in}/T_{out}$, in sharp contrast to the square root scaling that characterizes the NT limit, Eq.~(\ref{Case1}).

\begin{figure}[h]
\center
\includegraphics[width=\columnwidth]{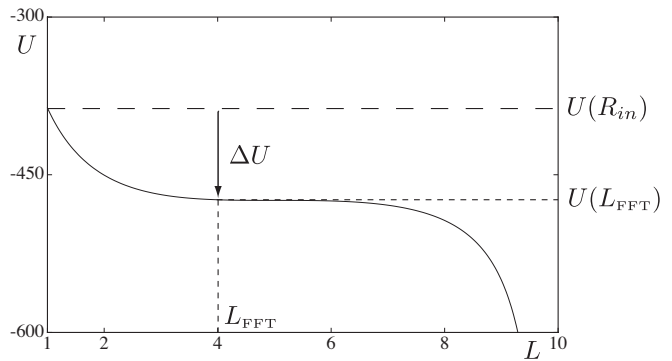}
\caption{The mechanical energy $U$ in the FFT limit (normalized by $ Y / (\Rin {T_{out}})^2$) as a function of the wrinkle extent $L$ (normalized by $\Rin$). Here we took $T_{in}/T_{out} \!=\! 8$, $R_{out}=10 \Rin$ and $\nu\!=\!1/3$. An inflection point exists at $L_{\mbox{\tiny FFT}}\!=\!T_{in}/2T_{out}=4$; the upper bound implies that this is the actual extent of the wrinkles. When reducing $\epsilon$ from $\epsilon_c$ in the NT  to the FFT regime ($\epsilon \! \ll\! 1$) the energy decreases by an amount $\Delta U$ from the upper dashed line (the energy of the Lam\'e solution) to the value at the inflection point, $U(L_{\mbox{\tiny FFT}})$.
}
\label{fig_2}
\end{figure}

Let us turn now to the number of wrinkles $m$ in the FFT limit $\epsilon \to 0$. While Eq.~(\ref{FFThoopstrain}) reveals that the product $m f(r)$ must remain finite, we do expect $|f|$ to become smaller and hence $m$ diverges as the FFT regime $\epsilon \! \ll \! 1$ is approached. This agrees with experimental observations \cite{Cerda01,Huang}, and is also a direct consequence of our theory \cite{Davidovitch10}. Physically, this is correlated with the fact that the dominant out-of-plane forces in the FFT regime are bending and compression in the azimuthal direction and stretching in the radial direction. To leading order in $\epsilon$, the $1^{st}$ FvK Eq.~(\ref{FvK22}) in the FFT regime is \cite{Davidovitch10}:
\begin{equation}
B \frac{m^4}{r^4} f = - \sigma_{\theta\theta} \frac{m^2}{r^2} f + \sigma_{rr} f^{''} \ .
\label{normalforceFFT}
\end{equation}
Balance of these three forces implies the hoop stress scaling $\sigma_{\theta\theta} \sim \sqrt{B T_{out}}/\Rin$ in contrast to the NT regime, where $\sigma_{\theta\theta}$ is a function of the exerted stresses $T_{in},T_{out}$ but is independent $B$, see Eq.~(\ref{Lame1}). One should notice that this result can be expressed as $\sigma_{\theta\theta}/T_{out} \sim \epsilon^{1/2}$, and reflects the collapse of compressive stress in the FFT regime that is the basis of our analysis. The same balance implies the FFT scaling law for the wrinkles number
\begin{equation}
m \sim k(\tau) (\Rin^2\, T_{out}/B)^{1/4} \sim \epsilon^{-1/4}
\label{wrinklesnumber}
\end{equation}
 for some $k(\tau)$, in sharp contrast to the NT scaling \eqref{NT:mscale}. This scaling law has already been predicted in \cite{Cerda01}, and is strongly supported by experimental observations on ultra thin-sheets, where $\epsilon$ is estimated to be below $10^{-6}$ \cite{Huang}.

In the experiments of \cite{Huang}, a very thin circular sheet is floating on water, subject to surface tension $T_{out} = \gamma$ at its perimeter $r=R_{out}$. A liquid drop is placed at the center, deforming the sheet beneath it, and exerting an in-plane tensile force $T_{in}$ at the contact line $r=\Rin$. The problem is thus analogous to the Lam\'e problem, but the determination of $T_{in}$ is a subtle problem that has not been resolved yet \cite{Vella}. Although our result~(\ref{wrinkleslength}) cannot be directly compared to the experiments of \cite{Huang}, it does provide a new answer to a puzzle raised by the empirical rule found there: $L_{\mbox{\tiny FFT}} \approx C_L \sqrt{Y/\gamma} \Rin$ with $C_L$  a numerical constant. The authors of \cite{Huang} assumed the NT scaling, Eq.~(\ref{Case1}), and concluded that their results indicate that $T_{in}$ is ``independent of surface tension, which is implausible". However, the FFT scaling, Eq.~(\ref{wrinkleslength}), shows that the empirical law is consistent with $T_{in} \sim \sqrt{Y\gamma}$, suggesting instead that the in-plane tension exerted by the drop at the contact line is affected ``equally" by the surface tension and the stretching modulus.

We conclude by making several comments on the nature of the FFT analysis that we have developed here. First, similarly to the NT analysis, we note that the wrinkle extent~(\ref{wrinkleslength}) is determined by minimizing the in-plane stretching energy which is much larger that the bending and out-of-plane stretching energies that determine the number of wrinkles~(\ref{wrinklesnumber}). This leads to the prediction that in wrinkling patterns the length is more robust than the number of wrinkles. Second, our theory of the FFT regime ($\epsilon \!\ll\!1$) is similar to a ``membrane limit" analysis which considers an elastic sheet with vanishing bending modulus (i.e. $\epsilon \!=\! 0$) \cite{Mansfield}. Our analysis, however, clarifies the singular nature of this limit: while the NT state can be obtained by regular expansion around the planar Lam\'e state \cite{Adams}, the FFT state is obtained by an asymptotic expansion around a limit in which $m$ diverges. Finally, one may wonder whether our theory enables an exact computation of the asymptotic number of wrinkles $m$ beyond the scaling law~(\ref{wrinklesnumber}). In order to do this, one must use Eqs.~(\ref{FFT1},\ref{ur1},\ref{FFThoopstrain}) to compute the bending energy $U_B\!=\! (B \!/\!4) \int_{\Rin}^L dr\, m^4 f^2(r)$ and the out-of-plane stretching energy $U_S\!=\!(1/4)\int_{\Rin}^L dr\, \sigma_{rr}(r) f'(r)^2$ as a function of $m$, and determine $m$ as the minimizer of these sub-dominant energies. One finds, however, a divergence of $f'(r)$ and $U_S$ as $L\to L_{\mbox{\tiny FFT}}$ for all $m$. Initially, this observation may lead one to doubt the validity of our results. However, careful thinking reveals the source of this divergence \cite{Davidovitch10}: our matching conditions at $r=L$ assume a direct transition from the planar state at $r> L$ to the fully-collapsed region at $r<L$. The divergence of $U_S$ is cured if one replaces the ``pointwise" matching by a narrow boundary layer
near $r \approx L$ in which the two regimes are smoothly matched. The energetic cost of this boundary layer is essential to find $m$, but requires a nontrivial calculation which is currently underway.

To summarize, we introduced an analysis of thin sheets far from buckling threshold. The central assumption of our theory, which should also be valid for more complicated set-ups, is the collapse of compressive stress in the wrinkled region. For the simple Lam\'e geometry we found a new scaling law for the wrinkle length and showed that it explains a puzzle arising from previous experiments~\cite{Huang}.
We anticipate that our work will lead to understanding of far-from-threshold wrinkling patterns under many different load configurations.




We acknowledge support by the Petroleum Research Fund of ACS (B.D.) and NSF-MRSEC on Polymers at UMass (R.S.). D.V.~is supported by an Oppenheimer Early Career Fellowship. M.A.B.~and E.C.~acknowledge the support of CNRS-Conicyt 2008. E.C.~thanks Fondecyt project 1095112 and Anillo Act 95.

\end{document}